\documentclass[sigconf,nonacm,pdfa]{acmart}

\newcommand\vldbdoi{10.14778/3675034.3675047}
\newcommand\vldbpages{2568 - 2575}
\newcommand\vldbvolume{17}
\newcommand\vldbissue{10}
\newcommand\vldbyear{2024}
\newcommand\vldbauthors{\authors}
\newcommand\vldbtitle{\shorttitle} 
\newcommand\vldbavailabilityurl{}
\newcommand\vldbpagestyle{empty} 

\usepackage{colorprofiles}
\usepackage[a-2b]{pdfx}

\usepackage{balance}
\usepackage{enumitem}

\begin{document}
\title{CXL and the Return of Scale-Up Database Engines}

\author{Alberto Lerner}
\affiliation{%
  \institution{eXascale Infolab \\
  University of Fribourg, Switzerland}
}
\email{alberto.lerner@unifr.ch}

\author{Gustavo Alonso}
\affiliation{%
  \institution{Systems Group, Department of Computer Science \\
  ETH Zurich, Switzerland}
}
\email{alonso@inf.ethz.ch}

\begin{abstract}
The trend toward specialized processing devices such as TPUs, DPUs, GPUs, and FPGAs has exposed the weaknesses of PCIe in interconnecting these devices and their hosts.
Several attempts have been proposed to improve, augment, or downright replace PCIe, and more recently, these efforts have converged into a standard called Compute Express Link (CXL).
CXL is already on version 2.0 in terms of commercial availability, but its potential to radically change the conventional server architecture has only just started to surface.
For example, CXL can increase the bandwidth and quantity of memory available to any single machine beyond what that machine can originally provide, most importantly, in a manner that is fully transparent to software applications.

We argue, however, that CXL can have a broader impact beyond memory expansion and deeply affect the architecture of data-intensive systems. 
In a nutshell, while the cloud favored \textit{scale-out} approaches that grew in capacity by adding full servers to a rack, CXL brings back \textit{scale-up} architectures that can grow by fine-tuning individual resources, all while transforming the rack into a large shared-memory machine.  
In this paper, we describe why such architectural transformations are now possible, how they benefit emerging heterogeneous hardware platforms for data-intensive systems, and the associated research challenges.
\end{abstract}

\maketitle

\pagestyle{\vldbpagestyle}
\begingroup\small\noindent\raggedright\textbf{PVLDB Reference Format:}\\
\vldbauthors. \vldbtitle. PVLDB, \vldbvolume(\vldbissue): \vldbpages, \vldbyear.\\
\href{https://doi.org/\vldbdoi}{doi:\vldbdoi}
\endgroup
\begingroup
\renewcommand\thefootnote{}\footnote{\noindent
This work is licensed under the Creative Commons BY-NC-ND 4.0 International License. Visit \url{https://creativecommons.org/licenses/by-nc-nd/4.0/} to view a copy of this license. For any use beyond those covered by this license, obtain permission by emailing \href{mailto:info@vldb.org}{info@vldb.org}. Copyright is held by the owner/author(s). Publication rights licensed to the VLDB Endowment. \\
\raggedright Proceedings of the VLDB Endowment, Vol. \vldbvolume, No. \vldbissue\ %
ISSN 2150-8097. \\
\href{https://doi.org/\vldbdoi}{doi:\vldbdoi} \\
}\addtocounter{footnote}{-1}\endgroup

\ifdefempty{\vldbavailabilityurl}{}{
\vspace{.3cm}
\begingroup\small\noindent\raggedright\textbf{PVLDB Artifact Availability:}\\
The source code, data, and/or other artifacts have been made available at \url{\vldbavailabilityurl}.
\endgroup
}

\section{Introduction} 
\label{sec:introduction}
The availability of elastic hardware in the cloud has enabled new, ever-demanding classes of applications to emerge. 
Many of these applications benefit from the acceleration provided by \textit{heterogeneous computing devices}, such as Smart NICs, FPGAs, GPUs, TPUs, and DPUs, to name the most common ones~\cite{Thompson21-decline}. 
These devices are often more powerful than the CPU for specific tasks~\cite{DBLP:conf/isca/ZhaoABGPOKPBLNL22,MaschiA23}, but they expose the fact that the most currently used system interconnect, the PCIe system~\cite{pcieSpecs}, is unsuitable for this purpose~\cite{DBLP:conf/sigmod/LutzBZRM20,DBLP:journals/pvldb/YuanL013}.
The bandwidth and latency incurred by moving data back and forth between CPU and devices erodes the devices' benefits and forces applications to batch data transfers and restructure themselves to hide the transfer latency.
Not surprisingly, recent devices are packed with memory to hold as much data as possible locally and avoid data movements through the PCIe system. 
The situation reached the point that in some computing devices, 75\% of the power and 80\% of their area is consumed by memory~\cite{dally2011power,dally2020domain}.

In response, hardware vendors have proposed a different approach to device integration.
Rather than copying data in and out of a device, the approach interconnects different accelerators and CPUs through distributed shared memory.
This gives CPUs and devices the ability to cache one another's memory coherently~\cite{sorin2011primer}.
Of course, the local memory size is still relevant, but with coherence, the application need not control how data moves; it simply accesses data as if it were local, and the added machinery efficiently relocates data as necessary.
Because coherency is mediated through hardware, which is designed to manipulate small data units, coherent transfers can achieve latency and bandwidth only moderately worse than what a CPU can achieve.
Several competing proposals ensued that allowed this type of memory-based integration to become public standards, such as OpenCAPI~\cite{opencapi}, CCIX~\cite{ccix}, and Gen-Z~\cite{genz}. These efforts have now converged into the \textit{Compute Express Link} standard (CXL)~\cite{cxl}.

CXL is a strict superset of PCIe, which thereby fosters easier adoption.
It is, however, much more than just an interconnect protocol.
It consists of three classes of interfaces. 
First, the \textit{I/O interface} increases the speed of PCIe without major changes to its semantics, i.e., it can still copy memory regions from one device to another in a non-coherent manner.
Second, the \textit{memory interface} allows a host CPU to coherently access the memory or storage of peripheral devices. 
Lastly, the \textit{cache interface} allows peripheral devices to coherently access and cache data from the host's memory.

The architectural change that the two latter interfaces represent was almost unimaginable just a few years ago.
Cache coherency protocols have existed for as long as caches have, but CPU manufacturers treated their coherency protocols as trade secrets.
For this reason, coherency could only be established across a vendor's CPU and its memory controllers.
With CXL, different CPUs can talk to different controllers, including those based on computing devices, because the protocol for doing so is now an open standard.

As big as this change is, it is not the only benefit CXL brings. 
In its latest version, the standard goes beyond the traditional role of an interconnect; it becomes an alternative to a rack-level networking fabric that is both more performant than current Ethernet-based systems as well as more expressive.
It offers additional features such as memory sharing and trusted memory enclaves, as opposed to just exposing a packet sending and receiving interface as in TCP/IP, or a remote memory access interface as in RDMA. 

In this paper, we focus on these capabilities of CXL and argue that they enable \textit{scale-up} (i.e., vertically scaled) database engines that significantly differ from the \textit{scale-out} (i.e., horizontally scaled) systems that have dominated the landscape in the last years due to the constraints imposed by cloud architectures. 

In what follows, we first provide a brief introduction to CXL~(Sec. \ref{sec:background}) and then present in detail three new architectural possibilities enabled by novel CXL features~(Sec. \ref{sec:single_server}), discussing the challenges that each brings.
We show that these architectures can further benefit from Near-Data Processing capabilities~(Sec.~\ref{sec:acceleration}) and by Heterogeneous Processing ones~(Sec.~\ref{sec:heterogeneous_architectures}).
Lastly, we also discuss related efforts~(Sec.~\ref{sec:related}) and close with our conclusions~(Sec.~\ref{sec:conclusion}).

\begin{figure*}[t]
\includegraphics[width=0.9\textwidth]{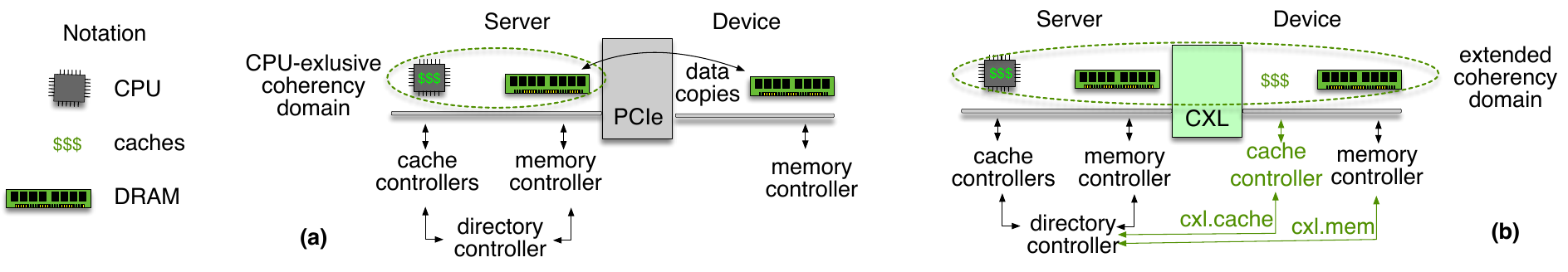}
\centering
\caption{
(a) Peripherals connected using PCIe cards are outside the coherency domain, even if they contain memory.
(b) CXL is a set of protocols that allow components on the peripheral to join the CPU coherency domain.
}
\label{fig:background}
\end{figure*}

\section{Background and Motivation}
\label{sec:background}
CXL is, in essence, a set of new protocols to connect hosts to peripheral devices.
It helps to start our discussion by revisiting how several cores in a given CPU are connected.

\subsection{Multicores and Memory Coherency}

A multicore server is a distributed system where CPU cores share the server's memory. 
Each core can cache data in small blocks, called \textit{cache lines}, occasionally duplicating that data. 
Data consistency issues may arise because two cores may wish to modify the same line simultaneously. 
\textit{Cache coherence} avoids the issues by maintaining two invariants:
(1) writes to the same memory location are serialized; and (2) every write is eventually made visible to all cores~\cite{sorin2011primer}.
Simply put, when a core intends to modify a cached line, it triggers a process known as cache invalidation, ensuring that only that copy of the line remains cached.

Cache coherence is handled by hardware components, usually via cache and memory controllers, as Figure~\ref{fig:background}(a) depicts. 
Until recently, the only entities that could cache lines were CPU cores, and the memory from which they could cache was limited to a server's available DRAM.
In this scenario, we say that the \textit{coherency domain} extends only to the CPU.

\subsection{Coherent and Non-Coherent Data Transfers}

Without CXL, data is copied between servers and devices through PCIe exchanges called \textit{transactions}.
PCIe transactions are not coherent because the device is outside the reach of the cache invalidation messages.
Data copies can quietly become stale.

CXL supports extended coherence domains by creating new types of transactions between servers and devices.
It does so in a backward-compatible way.
The PCIe transactions are wrapped in a sub-protocol called \texttt{cxl.io} and two other sub-protocols are added: \texttt{cxl.mem} and \texttt{cxl.cache}.
The server uses the former to communicate with the devices's memory controller as if it resided locally on its motherboard.
The device uses the latter sub-protocol to cache contents managed by the server's directory controller.
Figure~\ref{fig:background}(b) depicts the two new protocols. 
Because of the coherent sub-protocols, any device's memory can be seamlessly incorporated into the overall system, and the device can cache data that lives in the system's memory.
Devices are able to use one or both of the new sub-protocols. 
If they only use \texttt{cxl.cache} and do not contribute memory to the system, they are called \textit{Type 1} devices.
If they only use \texttt{cxl.mem} and do not cache memory from the host, they are called \textit{Type 3}.
Devices that use both are known as \textit{Type 2}.

\subsection{CXL Status and Availability}

CXL is a public consortium led by Intel, and it has had four major releases so far. 
The 1.1 version focuses on local memory expansion, allowing a server to access more memory than is available in its DIMM slots through locally attached CXL devices called \textit{memory expanders}.
Release 2.0 introduced basic forms of interconnects (i.e., CXL switches).
It allows memory expander devices to be installed in remote chassis and be shared by different servers.
As of this writing, two generations of Intel CPUs support CXL 1.1~\cite{sp4,sp5}, and a new generation supporting 2.0 was just released~\cite{xeon6}.
There are manufacturers offering Type 3 expanders, including some for CXL 2.0 (e.g., ~\cite{micronexpander,samsungexpander,samsungbox}).
The first CXL switch silicons and adapters are also becoming available~\cite{xconn,photowave}.
Most importantly, some established database vendors are evaluating these technologies~\cite{SAPElastic,SAPCXL2}.

CXL versions 3.0 and 3.1 support more sophisticated networking and additional memory-sharing modes, including peer-to-peer exchanges among devices.
These versions depend on improvements in PCIe, which will be accomplished in the already ratified PCIe Gen 6 and the Gen 7~\cite{pcieSpecs}. 
We expect the next generation of Intel and AMD CPUs to support PCIe 6 and the newest CXL versions.

\subsection{CXL Performance Characterization}

It is easy to look into CXL memory performance by comparing it to NUMA memory access.
Current conventional NUMA servers typically comprise two sockets, each containing a CPU and half of the system's DRAM. 
When a CXL memory expander is used, it effectively attaches more DRAM DIMMs to the system by creating an additional NUMA node, albeit one without any cores.
We note, however, that simply modeling CXL memory as NUMA can be inaccurate.
The number of memory controllers in an expander and, thus, available bandwidth varies from product to product.
Moreover, the type of memory used in the expander need not be the same as the one in the hosts.

Preliminary benchmarks by Microsoft, Meta, and Intel about CXL memory performance are already available (resp., ~\cite{TPP23,sun2023cxlcharacterization,Pond23}).
The perceived latency for CXL memory access is slightly higher than that for remote NUMA access, roughly in the 200--400 nanoseconds range.
As noted above, the bandwidth highly depends on the expander's characteristics.
It can be lower or higher than that of a memory controller found within a host NUMA node.
As just mentioned, nothing prevents an expander from using HBM instead of DDR memory.
We summarize the performance results of the three papers we mention next.

The micro-benchmarks from Intel investigate both latency and bandwidth~\cite{sun2023cxlcharacterization}. Regarding latency, executing a \texttt{load} instruction against a given type of CXL-attached memory can take just 35\% longer than the equivalent NUMA memory access.
Executing a \texttt{store} under the same conditions can present slightly lower but equivalent overheads. Regarding bandwidth, transfers from NUMA nodes can be 70\% efficient when considering only \texttt{load}s, compared to 46\% efficiency through a CXL interconnect.

Meta published results on the end-to-end impact of using CXL in applications \cite{TPP23}.
In that work, CXL memory stored cold pages, with the operating system swapping pages back and forth between the host DRAM and the CXL memory. 
In essence, CXL memory was used as an additional memory tier between the DRAM and storage. 
The results suggest that the bandwidth available from CXL memory is around 64~GB/s with latency only slightly larger than that of NUMA memory. 

Microsoft also performed an end-to-end study on the impact of CXL memory for their cloud environment~\cite{Pond23}.
While the results differ from workload to workload, the study found that under the expected latency increases, some 26\% of the 158 workloads studied showed less than 1\% performance penalty, and an additional 17\% showed less than 5\%.
For database workloads, specifically TPC-H, the overheads were highly query-dependent but were mostly below 20\%. 
This analysis already considers CXL switches and shared memory among a large number of machines in a rack.

\subsection{CXL as Networking}

Other studies suggest that CXL interconnects, the fabric carrying coherency traffic between servers and devices, can be significantly more efficient than RDMA-enabled networks~\cite{camelCXL,gouk2022disaggregatedmem,cxl}.
As seen above, the latency of CXL communications averages in the low hundreds of nanoseconds, while the fastest RDMA exchanges take a few microseconds---a difference of at least 2.5$\times$.
The advantages go beyond latency.
These studies also hint at bandwidth differences and partly attribute them to the fact that traditional network interface cards (NICs) underutilize the PCIe lanes they occupy. 
For instance, a 400~Gbps NIC (50~GB/s) uses 16 PCIe Gen5 lanes that, in the aggregate, can offer 64~GB/s~\cite{cx7}. 
This discrepancy means that over 20\% of the available PCIe bandwidth does not translate into network bandwidth.
CXL adapters utilize the full bandwidth.  

From the application point of view, replacing RDMA with CXL turns an application that is networking API-based into a many-core, shared-memory application.
This application can scale by launching additional threads and can access remote memory without incorporating networking APIs, such as Infiniband Verbs. 
Moreover, much of the memory access coordination is managed in hardware, through coherence, instead of by software. 
That is because CXL memory is accessed using the same familiar \texttt{load} and \texttt{store} instructions that applications use to access local memory.
The hardware makes the remote access look transparent to the software, including maintaining coherence.

Ultimately, CXL's advantages go beyond software simplicity and performance.
There are many functional aspects that RDMA simply cannot handle.
RDMA does not provide a path from the CPU to accelerators (computing devices) for instance; CXL does. 
RDMA does not provide a path between different server components, for instance, from a NIC to an SSD; CXL does. 
RDMA only enables access to disaggregated memory; CXL allows disaggregation of an entire rack without requiring that the devices support anything other than PCIe at the physical level.

\subsection{CXL Scalability and Fault Tolerance}

CXL imposes a limit on the number of devices that contribute and/or cache memory from each other.
Currently, a CXL coherence domain can support a \textit{diameter} of up to 4096 devices.
Arguably, this number of devices is amply sufficient to cover one rack---or even a small number of racks--- especially because not all the devices need to be coherent.
We note that such interconnection size limitations are not uncommon.
For instance, RDMA requires the underlying Ethernet network to be lossless, something that it is also difficult to do at a large scale~\cite{RDMAScale}.
We expect that the CXL limitations will diminish as it evolves, as is typical with emerging technologies.

Failures can, of course, occur on a disaggregated rack, whether using RDMA or CXL.
CXL does not add failure modes to those that are normally expected in a distributed system.
However, memory expansion through CXL brings at least two advantages in exceptional scenarios.
First, failure detection and propagation are built into the protocol through a set of mechanisms known as RAS (reliability, availability, and serviceability)~\cite{asteraRAS}. 
Because the hardware is conceived to identify and address failures, the reaction times in a CXL platform are likely faster than in a traditional distributed system.
Second, using CXL disaggregated memory instead of a remote machine's memory is a better scenario for failure probability due to the lower number of components and lower potential for load interference issues.

\begin{figure*}[t!]
\includegraphics[width=\textwidth]{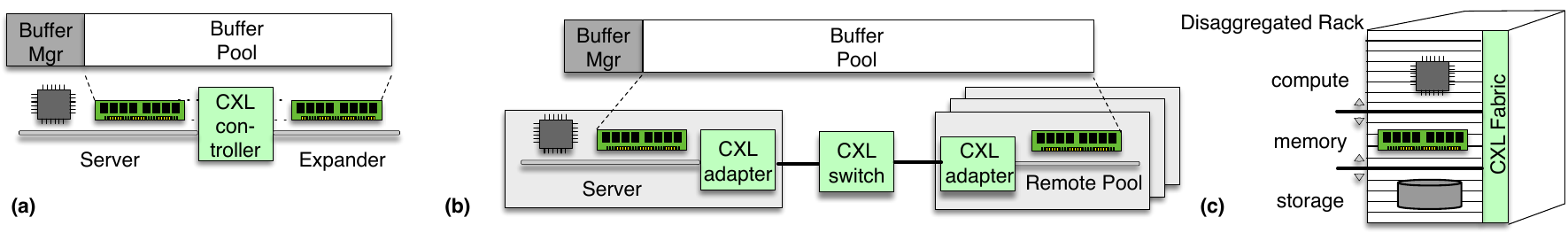}
\centering
\vspace{-18pt}
\caption{
CXL allows different memory expansion architectures: 
(a) local expansion and (b) far-memory pooling, and
(c) full-rack disaggregation.
The latter two are achieved through specialized CXL switches.}
\label{fig:disaggregation_progression}
\end{figure*}

\section{Shared-Memory Architectures}
\label{sec:single_server}
In this section, we will explore how CXL can expand a system's available memory in at least three ways. Figure~\ref{fig:disaggregation_progression} illustrates the resulting architectures. 
We will introduce each, discussing their implications, potential innovative applications, and the necessary research work to support them.

\subsection{Memory Expansion}

Database systems maintain a memory pool that all threads use to process queries or transactions called \textit{Buffer Cache} or \textit{Buffer Pool}.
In disk-based systems, the pool is an intermediate stage and cache between persistent storage and each thread's private space~\cite{diskDBARch}. 
In main-memory systems, the pool holds data in its entirety while still providing memory for query processing~\cite{memDBArch}.
The most straightforward way to expand a database system's memory is to have the pool also manage CXL memory, as Figure~\ref{fig:disaggregation_progression}(a) shows.
After adding an expansion card to a system, the pool can access CXL memory using the same instructions it uses for conventional memory, no API needed, while the hardware takes care of coherency.

However, a database system now has to match the different data placement options to the requirements for data access.
This mapping should be flexible enough to accommodate different types of memory expanders and dynamic enough to reflect a system's instantaneous workload.
Historically, this mapping was solved by a \textit{memory tiering} approach.
Hot pages are placed in fast-tier memory, and cold pages are in less-performing tiers.
A system such as an OS paging mechanism would track pages' \textit{temperatures} and make data placement decisions accordingly.
Some, like Meta mentioned above~\cite{TPP23}, advocate a similar approach for CXL memory.
Since it is long established that a database engine can better calculate the utility of keeping a page in a given memory tier than the OS~\cite{DBMin85}, we posit that existing I/O optimization and buffer management techniques that have been successful in other deep memory hierarchy scenarios can play a new role here~\cite{Memory-Bottleneck00,In-Memory-OLTP}. 

A ``killer app'' for this architecture may lie in the very memory diversity it supports.
In other words, the memory tiers can be carefully designed.
CXL memory expanders not only break the notorious memory wall~\cite{Memory-Bottleneck00} but also rid the system of inter-dependencies.
The expansion memory type need not match the memory type the current CPU/mother\-board supports. 
The expansion can contain regions with slower/cheaper or faster/more expensive memory than the CPU at the system architect's discretion, even enabling the recycling of DRAM from older generations, an aspect with the potential to result in significant cost savings and a reduction of the environmental footprint of computers. 
From a database perspective, an interesting configuration would place the transactional workload on the local DRAM and use CXL memory for the analytic part.
The data structures in the CXL memory could be specialized ones, such as data cubes, materialized tables, and denormalized tables, to cite a few.
Thanks to the data placement, the OLTP and OLAP data structures would not interfere with each other. 
Exercising this flexibility allows fine-tuning the memory characteristics (and cost) on which a given database system thrives. 

Efficiently expanding single-machine database systems with CXL memory raises several other important research questions:

\begin{itemize}[leftmargin=*,topsep=0pt,parsep=0pt]
\item Are memory expanders fast enough for OLTP or will they be suitable mainly for OLAP? Can they be used to perform both on the same machine and what are the implications?
\item Are the data structures kept in CXL memory the same as the ones that are successfully used in conventional memory?
If not, are there data structures suitable for the increased non-uniformity in memory accesses that CXL memory creates?
\item Should data structures span conventional and CXL memory? Could and should these structures be adaptable and offer dynamic data placement according to the access patterns?
\item Alternatively, should we forego designing data structures and use memory expanders as paging block devices for integration simplicity, or would specialized allocating techniques benefit from byte-addressable memory?
\end{itemize}

\subsection{Memory Pooling}

Unfortunately, increasing individual servers' memory with expand\-ers in a cloud setting can create a phenomenon called \textit{stranded memory}.
The memory physically present in one server may be underutilized by the systems running on it.
Hyperscalers have reported that, as memory is one of the most expensive components in data centers, stranding is a major source of inefficiencies~\cite{Pond23}.

Recently, researchers have explored ways to reassign such stran\-ded memory~\cite{Aguilera23,Redy21,Farview22,Stratos22}. 
The ideas have led to the notions of \textit{far memory}, \textit{remote memory}, and \textit{disaggregated memory}. 
Far memory is a term typically used to refer to any generic configuration where the additional DRAM is not local. 
Remote memory is also far but generally refers to utilizing unused memory of other machines. 
Disaggregated memory is not allocated to any machine but is available for any servers to use. 
In most cases, the connection between server and memory is made through RDMA and the remote memory is treated as a block device with a paging mechanism to move data back and forth between the host and the far memory. 

With CXL, a new architecture is possible that uses large, independent memory pools that can be carved into smaller pieces.
Each server can access one or more of these pieces through CXL switches, and a single such memory pool can serve multiple servers simultaneously this way. Figure~\ref{fig:disaggregation_progression}(b) depicts such a scenario.
As discussed above, one of CXL's main innovations is its ability to assign a portion of a pool to a server without requiring its applications to access it through a non-coherent networking API.
Direct, coherent memory access is far more efficient than what can be achieved in a distributed system that requires an external memory access API and leaves coherency to the application. 
Moreover, by cascading multiple CXL switches, CPUs on different machines can access a central memory expander containing a pool of memory available to all machines in a rack, 
very much like disaggregated storage is available in the cloud. 

We posit that this architecture's killer app is making database systems better cloud denizens.
Traditional databases assume that large amounts of data are loaded into a VM memory because the database performance depends on limiting I/O. 
The resulting VM is not elastic because databases do not cope well with dynamically varying the size of their buffer pools.
For the same reason, database systems also miss the benefits of serverless technologies. 
The cost of quickly loading and unloading the state is too high.

Disaggregated memory implemented through CXL allows for placing the buffer pool on the disaggregated memory and using the local DRAM for query processing. 
If more query processing capacity is needed, new engines can be spawned and connected to the disaggregated memory so that these engines are immediately ready to run queries, as there is no need to warm up the database. 

The additional latency of CXL memory plays only a minor role in databases~\cite{Pond23} and is a good trade-off for far more elasticity than is feasible with engines where the data resides in local memory. Similarly, a database engine can be easily migrated when the buffer pool is in disaggregated memory. If the data structures and state of the engine itself are maintained in disaggregated memory, then migrating the entire engine to another machine becomes a far simpler operation. 

From a research perspective, more interesting questions arise:

\begin{itemize}[leftmargin=*,topsep=0pt,parsep=0pt]
\item 
Implementing cloud-friendly features requires rethinking the internal database architecture to remove the assumption that everything is in local memory. 
Extensive experimentation of what needs to be local and what can be placed in disaggregated memory will be required to determine which part of the engine can tolerate the additional latency of CXL memory.
We discussed some of these architectural changes elsewhere~\cite{dataflows}.
\item Assuming database systems become more elastic using the above techniques, should the granularity to consider be at the entire engine level, or can the elasticity be pushed down to the level of threads running queries? 
\item Threads running queries could be moved from machine to machine by keeping their state and working space in disaggregated memory.
Alternatively, they can be created as the workload dictates. 
How would an engine operate under such a dynamically changing multiprogramming level?
\end{itemize}

\subsection{Memory Sharing}

We have seen how CXL can increase the memory availability and flexibility of single-server database systems.
The next question to arise is whether CXL can also benefit larger database systems.

Traditionally, these systems scale by adopting distributed database techniques~\cite{aguilar2020polaris,oracleexadata,AmazonAurora}. 
Simply applying the known scalability techniques into a CXL environment is unlikely to succeed for numerous reasons: to cite only a few, cache invalidation now crosses machines; updates imply more-onerous distributed, locks; and memory is wasted as the same data is copied into the local buffer cache of several machines. 
Some of these issues were attenuated in the past through a mix of replication and sharding, often using RDMA to reduce the network overhead \cite{RDMA-Database16,barthels192PL,yacine2018replrdma,zamanian2019replrdma}. 
However, these techniques invariably impose a notoriously fragile balance between consistency, symmetry in the system (e.g., with read-only copies), and data placement to minimize data movement, thereby 
limiting architectural freedom (e.g., \cite{Aurora-Distribution18}).

In contrast, CXL can integrate multiple servers and closely-placed large resource pooling modules without any of the above restrictions.
The pooling modules can contain homogeneous DRAM devices but also devices with different mixes of volatile and non-volatile memory (e.g.,~\cite{samsungCXLStorage}), creating storage pooling modules.
Advanced CXL features such as \textit{Global Integrated Memory} (GIM) and \textit{Global Fabric-Attached Memory Devide} (GFAM), allow servers to share resources in the pooling modules collectively~\cite{cxl31}.
Simply put, the entire rack becomes a single shared-memory machine.
Figure~\ref{fig:disaggregation_progression}(c) depicts this scenario within one rack, but we believe the same features could also support spanning a small number of racks.

A killer app for this architecture is a larger database system in a truly scaled-up manner.
Each thread in this database could run on a separate compute module but share the same memory and storage map with threads running in another computer module.

A fully disaggregated system like the one CXL enables raises yet unexplored scalability questions, such as:

\begin{itemize}[leftmargin=*,topsep=0pt,parsep=0pt]
\item Hashing and sorting are at the core of most relational data processing, but it is not obvious how they would work at a rack-level scale.
It is possible that accepted wisdom regarding when to use each one will change in a scale-out architecture.
\item Given that each core can now access one to two orders of magnitude more memory than before, are the data structures we use to organize and index the data still effective at these new scales? Since the invalidation messages can now dominate access time, how is the \textit{coherency traffic} generated by a typical data structure? 
\item Assuming we now have the freedom to engage a tremendous amount of resources to solve individual query operators, how do we schedule the machine resources across competing queries?
\end{itemize}

\vspace{0.5em}
\noindent
Rack-level integration is arguably the most impactful change enabled by CXL.
It moves the memory unification effort to the hardware and thereby liberates the database system from managing coherence across servers. 
Moreover, CXL can attain bandwidth levels larger than those attained by today's networks and with far lower latency.  
We expect these possibilities alone to radically change the design of scalable databases and data processing engines, but there are two other profound possibilities.
We describe the first in the next section and the second in the subsequent one.

\section{Near-Data Processing}
\label{sec:acceleration}
CXL enables another important feature that applies to all the architectures we just described.
To understand this feature, it helps to examine how a CXL device is built.
A component in the device called coherency controller implements the CXL protocol.
In a memory expander, this controller wraps existing DRAM DIMMs and intercepts all the requests to memory.
Similarly, in a device that caches CXL memory, the controller maintains the cache state and responds to invalidation traffic.
Put differently, the CXL controller is an independent entity from the memory it manages. 

\begin{figure}[t]
\includegraphics[width=0.8\columnwidth]{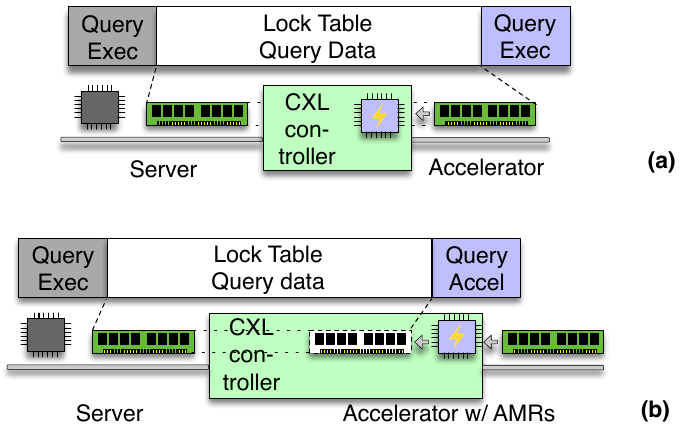}
\centering
\vspace{-8pt}
\caption{
Near-data processing under CXL:
(a) A processor or FPGA managing the expanded memory can be co-opted to execute a portion of a query. 
(b) A unique opportunity for acceleration exists through \textit{active memory regions} (AMR).
}
\vspace{-12pt}
\label{fig:acceleration}
\end{figure}

This independence opens a tremendous opportunity: 
The controller can be co-opted to perform computations over the data it transports.
This ability is known as \textit{near-data processing} (NDP).
Figure~\ref{fig:acceleration}(a) shows a controller capable of performing query processing on the device's side of the setup.

One may argue that NDP is hardly new.
There are examples where a small processor is placed near memory \cite{Accorda19} or in disaggregated memory \cite{Farview22} to accelerate query processing in research prototypes.
Some products, such as the Oracle's retired SPARC S7, featured Data Analytics Accelerators (DAX) to support the offloading of basic relational operators~\cite{Sparc-S7}.
However, in traditional NDP, query processing work is divided between the CPU and the accelerator because the accelerators do not support coherence.
If one side changes the data, the other will not see it.

In contrast, CXL-supported NDP allows both devices to work in parallel, not only because both can access data but also because the database system lock table can be shared, allowing each side to coordinate with the other.
Being able to parallelize work that has been offloaded allows us to design
a new generation of near-data accelerators that are more flexible
and, most likely, faster.

A killer app for CXL-supported NDP is a new take on virtualizing memory. 
Since the controller intercepts all CXL traffic, it can implement an \textit{active memory region} that is not backed by DRAM.
Instead, the region corresponds to a computation that is executed when its memory addresses are accessed.
If the computation is a streaming  one, the results need not be materialized and could be fed to the application as it reads memory.
Figure~\ref{fig:acceleration}(b) illustrates such a use case. 
In databases, this can be used for on-the-fly implementations of view materialization and data transpositions~\cite{Farview22} or better integration databases and smart storage systems~\cite{lee2024dbkernels}.

CXL acceleration is not without challenges, some of which we list below:
\begin{itemize}[leftmargin=*,topsep=0pt,parsep=0pt]
\item What portions of query processing should be done near the data?
Some examples, such as compression and decompression, encryption and decryption, selection, projection, and filtering with LIKE predicates, have brought substantial advantages in practice \cite{AWSaqua,Accorda19,Farview22,Chiosa22}.
We believe other opportunities still exist.
\item It is not inconceivable that CXL can support improving other fundamental mechanisms that are central to OLTP, such as collective communication, locking, timestamps, and dynamic data relocation across the system, among others.
\item 
Many near-data processing accelerators have memory hierarchies as well. 
Should we treat them as local memory or use them as CXL memory, and in which proportion?
\end{itemize}

\section{Heterogeneous Architectures}
\label{sec:heterogeneous_architectures}
We have depicted in Figure~\ref{fig:disaggregation_progression}(c) how to use CXL to connect pools of different resources, such as memory or storage. 
The same techniques allow other types of resources, such as FPGAs, GPUs, TPUs, and DPUs, to be similarly pooled and integrated into a rack-scale computer.
For instance, as we just discussed, an accelerated CXL controller may offer services based on those resources through an active memory range.

This composability possibility opens a research field of its own: 
\begin{itemize}[leftmargin=*,topsep=0pt,parsep=0pt]
\item 
What should a heterogeneous machine look like now that different processing devices can be independently pooled? 
How should we select devices and balance resources across them?
\item
Can we and should we build machines tailored to specific workloads?
For instance, machine learning (ML) is taxing database engines because the data often has to be taken out of the database to run it through ML tools. 
A heterogeneous architecture that seamlessly integrates CPUs and GPUs makes it possible to implement ML operators directly on the database engine.  
\end{itemize}

\section{Related Efforts}
\label{sec:related}
CXL is the result of consolidating several projects that came before it, most notably CCIX~\cite{ccix}, GenZ~\cite{genz}, and OpenCAPI~\cite{opencapi}.
An overwhelming number of institutions and companies are working together to advance the protocol~\cite{cxlmembers}.
The consolidation, however, has not been complete.
Some proprietary memory coherency interconnects still exist, mainly involving GPGPU vendors.
AMD supports an interconnect called Infinity Architecture~\cite{infinityfabric}, and  NVidia has NVLink~\cite{nvlink} and NVlink-C2C \cite{NVLink-C2C}.

The argument for developing specialized interconnects for GPUs is that they can provide higher bandwidths than what is currently available with PCIe.
While these arguments may have been valid with older versions of PCIe, the latter standard has been upgraded at an unprecedented pace.
PCIe Gen 7, expected to be available in 2025, will support 128MT/s per lane or 242GB/s in a $\times$16 card~\cite{pci7}.
Even if proprietary interconnects were to keep improving their bandwidth, they remain proprietary.
The reach of CXL reach is much wider than that of GPUs and, for that, it is a much more capable integration technology.

As with any new technology, some works have pointed to perceived CXL disadvantages, such as additional equipment cost or potential increase in software complexity~\cite{againstCXL}.
For database systems, however, CXL can eliminate the memory bottleneck that has plagued them for a while---something that can be done with minimal software
changes---and create possibilities for innovative architectures that were not feasible before. 
Any equipment cost that this might entail would be more than offset by the improving memory utilization, let alone other parallelism and acceleration opportunities that these new architectures can bring.

\section{Conclusion}
\label{sec:conclusion}
In this paper, we argued that CXL will reverse at least two decades of investments in scale-out database systems.
Through seamless integration of devices' and hosts' memory, CXL allows database systems to grow via scaling up rather than scaling out, turning what is now complex distributed system development into centralized system development while simplifying the development of heterogeneous systems in the process.
Given the magnitude of these possibilities, we expect CXL to catalyze the creation of a new generation of database systems with unprecedented scalability, efficiency, and ease of programming.

\section*{Acknowledgements}
This work has received funding from the Swiss State Secretariat for Education (SERI) in the context of the SmartEdge EU project (grant agreement No. 101092908).

\balance

\bibliographystyle{ACM-Reference-Format}
\bibliography{references}

\end{document}